\documentclass[aps,twocolumn,pre,showpacs,balancelastpage,amssymb,groupedaddress]{revtex4}
\usepackage{graphicx}
\usepackage{amsmath}

\newcommand{\tr}{\operatorname{tr}}
\newcommand{\lan}{\langle}
\newcommand{\ran}{\rangle}
\newcommand{\la}{\left\langle}
\newcommand{\ra}{\right\rangle}
\newcommand{\lv}{\left|}
\newcommand{\rv}{\right|}
\newcommand{\ei}{\mathrm{e}}
\newcommand{\bc}{\begin{center}}
\newcommand{\ec}{\end{center}}
\newcommand{\be}{\begin{equation}}
\newcommand{\ee}{\end{equation}}
\newcommand{\bq}{\begin{eqnarray}}
\newcommand{\eq}{\end{eqnarray}}

\begin{document}
\title{Quantifying the decay of quantum properties in single-mode states}
\author{L. A. M. Souza$^1$ \footnote{e-mail:
lamsouza@fisica.ufmg.br}, M. C. Nemes$^1$ , M. Fran\c{c}a Santos$^1$
and J. G. Peixoto de Faria$^2$}
\affiliation{$^1$Departamento de F\'{i}sica, Instituto de
Ci\^{e}ncias Exatas, Universidade Federal de Minas Gerais, CP 702,
CEP 30161-970, Belo Horizonte, Minas Gerais, Brasil; \\ $^2$
Departamento de F\'{i}sica e Matem\'{a}tica, Centro Federal de
Educa\c{c}\~{a}o Tecnol\'{o}gica de Minas Gerais, 30510-000, Belo
Horizonte, MG, Brasil.}

\begin{abstract}
The dissipative dynamics of Gaussian squeezed states (GSS) and
coherent superposition states (CSS) are analytically obtained and
compared. Time scales for sustaining different quantum properties
such as squeezing, negativity of the Wigner function or photon
number distribution are calculated. Some of these characteristic
times also depend on initial conditions. For example, in the
particular case of squeezing, we find that while the squeezing of
CSS is only visible for small enough values of the field
intensity, in GSS it is independent of this quantity, which may be
experimentally advantageous. The asymptotic dynamics however is
quite similar as revealed by the time evolution of the fidelity
between states of the two classes.

\end{abstract}


\maketitle

\section{Introduction}

The recent rapid development of quantum information theory has
largely stimulated research on nonclassical states of light. A
particularly promising approach consists in processing quantum
information with continuous variables \cite{braustein1}, where the
information is encoded into two conjugate quadratures of the
quantized mode of the optical field. Natural candidates for these
applications are Gaussian squeezed states (for a review of
experiments with squeezed light see \cite{bachor}) and
superpositions of two coherent states \cite{knightcoherent}. It is
imperative, therefore, to understand the quantum-classical limit
for these states which ultimately decides whether a particular one
can be used to enhance processing power in a quantum computer.

Classicality cannot be decided on the measurement of a single observable, i.e. a
classical description may explain some behavior and fail to
explain another. In particular, for continuous variable states,
studying the quantum-classical
limit implicates in analyzing quantum properties
(see
\cite{knight1,mandel,adam,buzek,barranco,knightsqueeze,buzek1,schleich,hong,mandel1,stoler,marian1,marian3}
and references therein) such as squeezing, oscillations in the
photon number distribution and sub-Poissonian statistics
\cite{mandel4}. This analysis can be statical or dynamical. For example, when comparing
two different classes of quantum states, let's say Gaussian squeezed states and superpositions of two
coherent states, one can ask how squeezed each one can be given a certain amount of invested energy or
how fast they lose this squeezing when coupled to an external reservoir, which is always present.
In fact, several recent experiments have shown that all quantum features of
light are sensitive to dissipation. Squeezing, oscillations in photon number distribution and interference
effects are dramatically reduced when the physical system is
coupled to a macroscopic environment. There is a vast literature
on both cases (see \cite{italianos} and references therein for a
review).

The question we address in this contribution is: how fast
do the two classes of initial states (CSS or GSS) loose their
potentiality to exhibit each one of the quantum properties they
have in common and/or others?

In other words, we analytically derive, according to
the model presented, characteristic time scales for different
quantum properties in both classes of states: GSS and CSS. In
particular, we show that although these classes of states share
some quantum features as e.g. squeezing and oscillations in photon
number distribution, their time scales are both quantitatively and
qualitatively different. For superposition states, there is a well
known time scale related to the time it takes for the
corresponding Wigner functions to become completely positive. This
is called decoherence time. It means that, in this time scale,
interference effects become unobservable. We show that the time
scale for observing squeezing effects or oscillations in photon
number distribution is \emph{smaller}~ than the decoherence time
and this limitation is due to the dependence of these times on the
intensity of the coherent fields. For GSS, however, the situation
drastically changes. The characteristic times for the observation
of the same properties are \emph{independent}~ of the intensity,
an effect obtained numerically in reference \cite{numerico}. In
this contribution, for the sake of comparison, we consider states
of both classes with comparable characteristic times. This allows
us to show that the dynamics leading to the asymptotic stationary
state is very similar, so that all interesting physics is
contained in the transient times. Moreover, for the GSS considered
it is possible to obtain an upper limit on the initial thermal
excitations so that quantum properties may still be visible
\cite{caroleeu}.

This work is divided as follows: we study some quantum properties
of coherent superposition states (section \ref{superposition}) and
of displaced, squeezed, thermal states (section \ref{DSTS}). The
quantum properties are, in both cases and respectively: squeezing
and interference (subsections \ref{catsquee}-\ref{gausssquee}),
oscillations in photon number distribution (subsections
\ref{catosci}-\ref{gaussosci}) and the von Neumann entropy
(subsections \ref{catent}-\ref{gaussent}). Besides the qualitative
comparisons made in the cited sections, in section \ref{secfid} we
present an analytical expression for the fidelity between the
superposition states and the GSS as a function of time. In section
\ref{conclusions} we briefly summarize this work and present the
conclusions.

\section{Superposition States: dissipative dynamics of quantum
properties} \label{superposition}

The class of initial states here considered can be written as pure
superposition states of the form \be \label{psi} \lv\psi\ra =
\frac{1}{N} \Big{(} \lv\beta_0\ra + \ei^{i \theta} \lv-\beta_0\ra
\Big{)}, \ee where \be \label{normalizacao} N = \sqrt{2(1+ \ei^{-2
|\beta_0|^2}\cos \theta )} \ee and $\lv\beta_0\ra$ stands for a
coherent state. If $\theta = 0 (\pi)$ we will have an even (odd)
coherent superposition state (which we shall, from now on call
even (odd) superposition states). These states can be produced both
in cavity QED \cite{haroche1} and propagating pulses \cite{pulses}
and have been utilized mainly to study the effects of decoherence and quantum-classical transition.

The dissipative dynamics we have in mind is the one well known
from quantum optics \be \label{master} \dot{\rho} = \mathcal{L}
\rho, \ee with \bq \label{liouvilliano1} \mathcal{L}\cdot &=& - i \omega [a^\dagger a ,
\cdot] + k (\bar{n}_B + 1) (2 a \cdot a^\dagger - a^\dagger a
\cdot - \cdot a^\dagger a)  \nonumber
\\ && + k ~ \bar{n}_B (2 a^\dagger \cdot a - a a^\dagger \cdot -
\cdot a a^\dagger) , \eq where $\omega$ is the angular frequency
of the unitary evolution, $\bar{n}_B$ is the mean number of
excitations in the environment, $k$ is the system-environment
coupling constant and $a$($a^{\dagger}$) is the annihilation
(creation) operator of one excitation in the field mode. In this
work we use units such that $\hbar = 1$.

The solution for the zero temperature case (also well known) is given by \bq
\rho_A(t) &=& \frac{1}{N^2} \Big{\{} \lv\beta(t)\ra
\la\beta(t)\rv + \lv-\beta(t)\ra \la-\beta(t)\rv + f(\beta_0, t) \times
\nonumber \\ \label{rhobeta}&& \times (\ei^{i \theta} \lv\beta(t)\ra \la-\beta(t)\rv+ \ei^{-i \theta}
\lv-\beta(t)\ra \la\beta(t)\rv) \Big{\}}, \eq where $N$ is given in
equation (\ref{normalizacao}), \be \label{ft} f(\beta_0,t) =
\ei^{-2(|\beta_0|^2-|\beta(t)|^2)} \ee and \be \label{betat}
\beta(t) = \beta_0 \ei^{-(i \omega + k)t}. \ee Note that the
stationary solution is a pure Gaussian state \be
\rho_A(t\rightarrow \infty) = \lv0\ra \la0\rv, \ee where $\lv0\ra$
is such that $a \lv0\ra = 0$. In this section we will only consider
zero temperature for simplicity.

The density operator \eqref{rhobeta} has the following eigenvectors
\cite{ze} \bq \lv e(t)\ra &=& \frac{1}{N_e(t)} \left[\lv\beta(t)\ra +
\lv-\beta(t)\ra\right] \nonumber, \\ && \\ \lv o(t)\ra &=& \frac{1}{N_o(t)}
\left[\lv\beta(t)\ra - \lv-\beta(t)\ra \right] \nonumber\eq whose eigenvalues are
given by \bq p_e(t) &=& \frac{1}{N^2} (1+\ei^{-2|\beta(t)|^2})
(1+\ei^{-2(|\beta_0|^2-|\beta(t)|^2)}\cos\theta)  \nonumber, \\
\label{autovalores} && \\ p_o(t) &=& \frac{1}{N^2}
(1-\ei^{-2|\beta(t)|^2})(1-\ei^{-2(|\beta_0|^2-|\beta(t)|^2)}\cos\theta),\nonumber
\eq where \bq N_e(t) &=& \sqrt{2(1+\ei^{-2|\beta(t)|^2})} \nonumber
\\ && \\ N_o(t) &=& \sqrt{2(1-\ei^{-2|\beta(t)|^2})}\nonumber. \eq
The subscripts $e$ and $o$ stand for even and odd superpositions
states respectively.

\subsection{Interference and Squeezing} \label{catsquee}

The potentiality of superposition states to produce
interference is encoded in the negative parts of their Wigner
functions. In the present case, the time evolution of the Wigner
function for states of the form \eqref{psi} is given by \bq
W(\lambda, \lambda^*,t) &=& \frac{4}{N^2} \left\{ \ei^{-2
|\beta(t)|^2 -2 |\lambda|^2}
\cosh\left[4 |\beta(t)|^2 \Re \left(\frac{\lambda}{\beta(t)}\right)\right] \right.\nonumber
\\ && \left.+  f(\beta_0,t) \ei^{-2 |\lambda|^2} \cos\left[\theta
-2|\beta(t)|^2 \Im \left(\frac{\lambda}{\beta(t)}\right)\right] \right\}. \nonumber \\
\eq The above expression has been obtained using the relation
$W(\lambda, \lambda^*) = 2 ~ \tr[\rho_A \mathcal{D}(\lambda)
(-1)^{\widehat{n}} \mathcal{D}^{-1}(\lambda)]$, where
$\mathcal{D}(\lambda)$ is the displacement operator,
$(-1)^{\widehat{n}}$ is the parity operator, $\lambda =
\sqrt{\frac{\omega}{2}}(x+i p)$ and the symbols $\Re$, $\Im$ stand
for real and imaginary parts respectively.

\begin{figure}[!h]
\begin{center}
\includegraphics [width = 6cm,height = 5.5cm, angle=0] {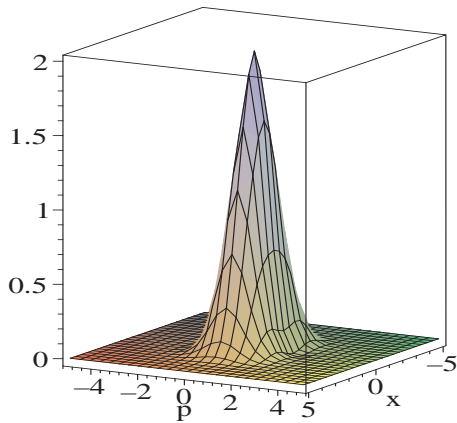}
\end{center}
\caption{Wigner function for the even superposition state ($\theta
= 0$). Parameters: $\beta_0 = 0.8$, $\omega = 1$, $k = 0.1$, $t =
0$.} \label{figure1}
\end{figure}

\begin{figure} [!h]
\begin{center} 
\includegraphics [width = 6cm,height = 5.5cm, angle=0] {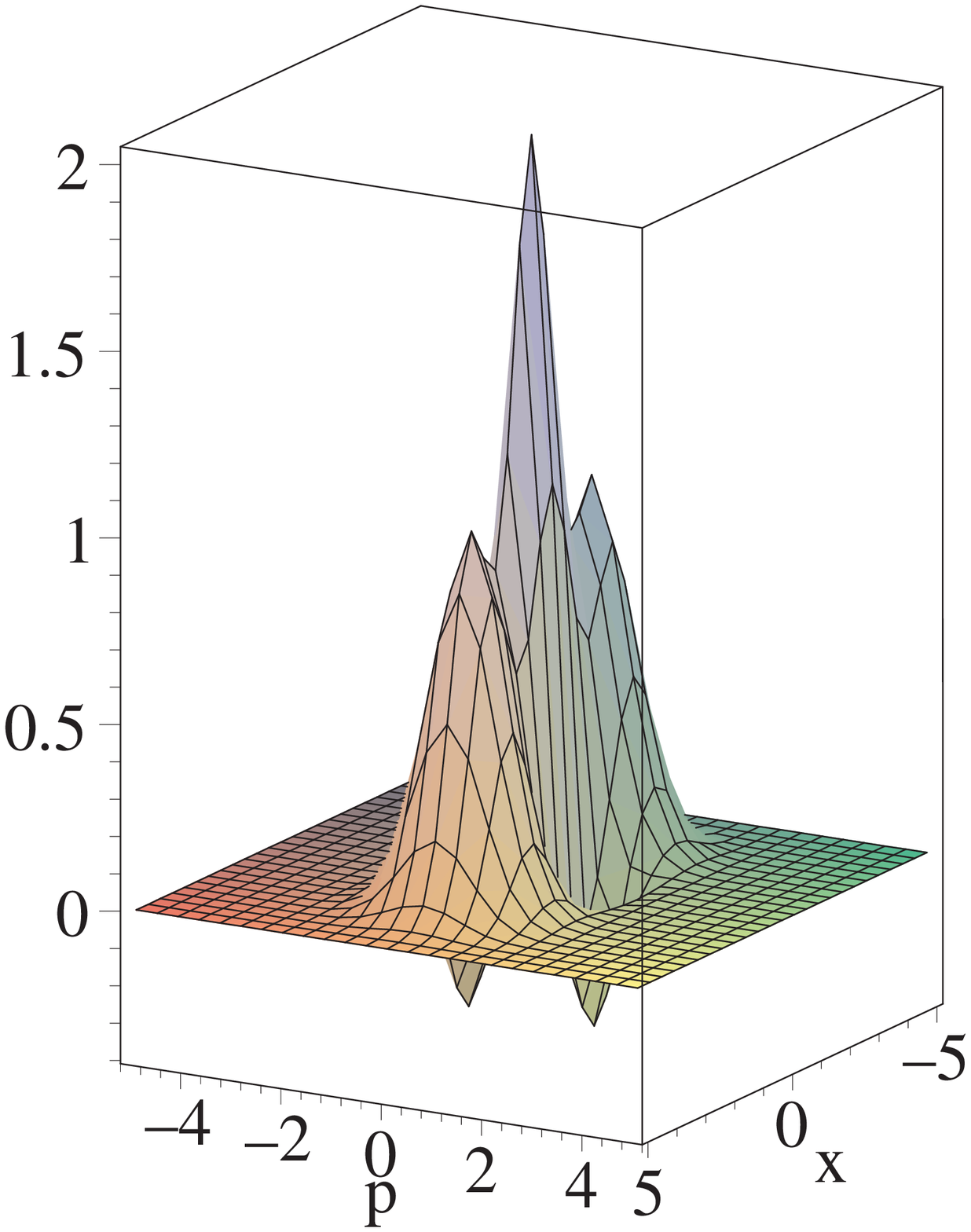}
\end{center}
\caption{Wigner function for the even superposition state ($\theta
= 0$). Parameters: $\beta_0 = 1.5$, $\omega = 1$, $k = 0.1$, $t =
0$.} \label{figure2}
\end{figure}

\begin{figure} [!h]
\begin{center} 
\includegraphics [width = 6cm,height = 5.5cm, angle=0] {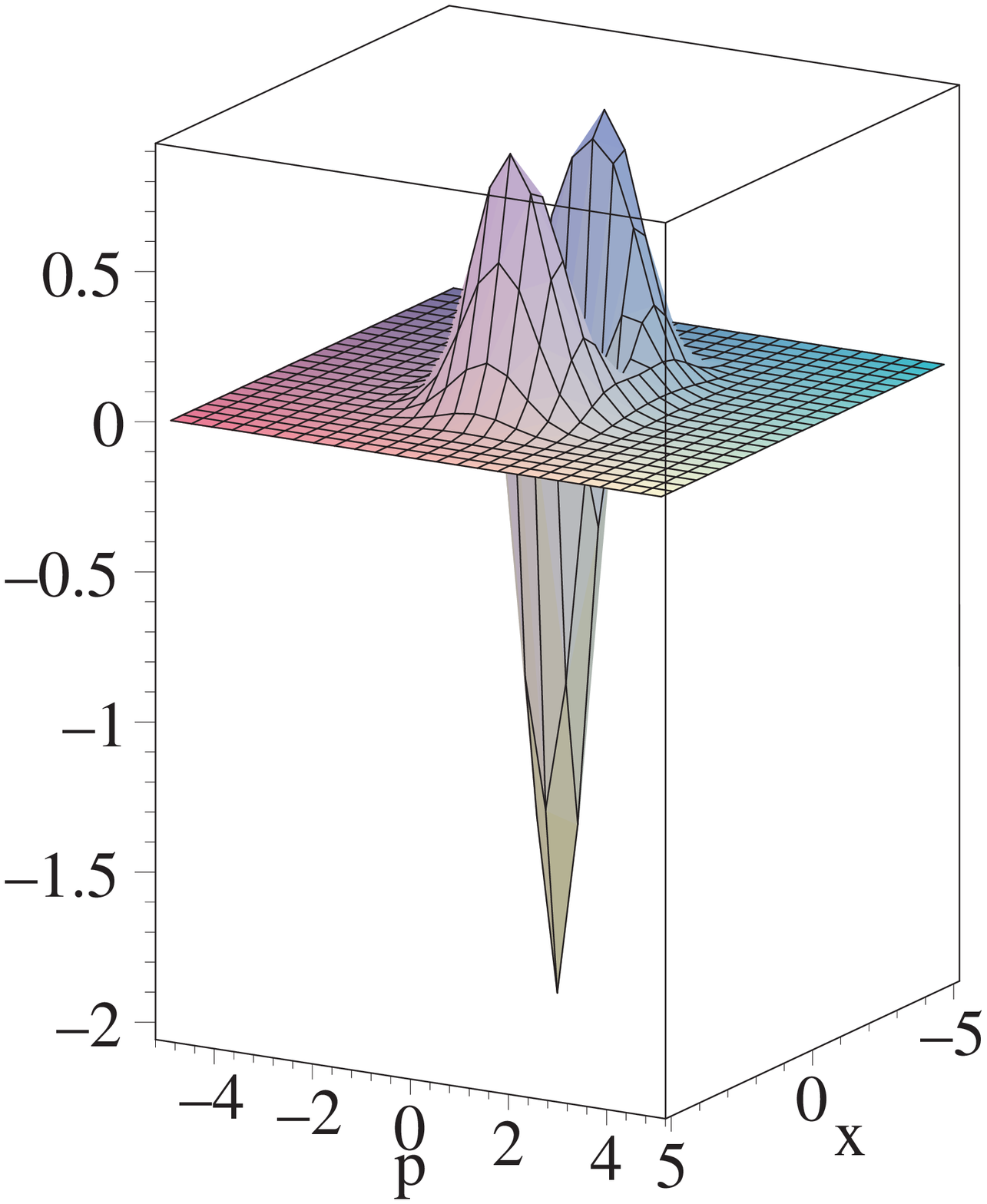}
\end{center}
\caption{Wigner function for the even superposition state ($\theta
= \pi$). Parameters: $\beta_0 = 0.8$, $\omega = 1$, $k = 0.1$, $t
= 0$.} \label{figure3}
\end{figure}

\begin{figure} [!h]
\begin{center}
\includegraphics [width = 6cm,height = 5.5cm, angle=0] {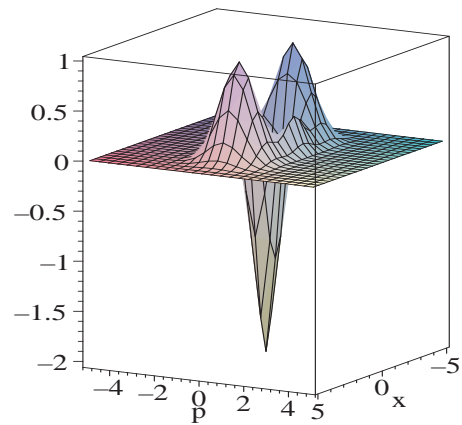}
\end{center}
\caption{Wigner function for the even superposition state ($\theta
= \pi$). Parameters: $\beta_0 = 1.5$, $\omega = 1$, $k = 0.1$, $t
= 0$.} \label{figure4}
\end{figure}

Some results for the Wigner function (WF) at time zero ($t = 0$)
of the even and odd coherent superposition states are shown
in figures \eqref{figure1} - \eqref{figure4}, for different values
of $\beta_0$.

The negative part of the WF for the same value of $\beta_0$ is
systematically larger for odd superposition states. However it
disappears simultaneously in the well known decoherence time $\tau
= \left(4 |\beta_0|^2 k\right)^{-1}$. When this time is reached,
there are no quantum effects which remain, neither squeezing nor
oscillating photon number distribution, as we shall show shortly.
These phenomena however possess different time scales than that of
decoherence. In figure \eqref{figure1} it is clear (at least
qualitatively) that the WF of a even CSS, with small values of
$\beta_0$, is quasi-Gaussian, the crucial difference is that it
shows interference (the negative parts). When we increase the
value of $\beta_0$, the negative parts of the even CSS are clear.
The odd CSS always has a negative value of Wigner function,
particularly in the point $x=p=0$.

\begin{figure} [!h]
\begin{center} 
\includegraphics [bb=0 30 217 183,scale=1.0] {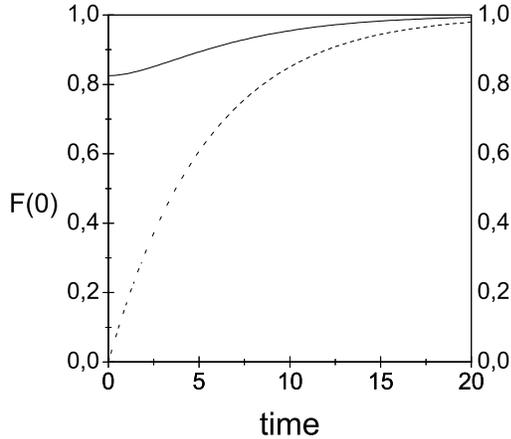}
\end{center}
\caption{Fidelity between the vacuum state and even ($\theta = 0$,
solid line) and odd ($\theta = \pi$, dashed line) superposition
states. Parameters: $\beta_0 = 0.8$, $\omega = 1$, $k = 0.1$. The
time scale is: $\mbox{time} = k t$.} \label{fideli}
\end{figure}

An interesting physically appealing way to understand
the dynamics of the WF for the even and odd superposition states
is as follows: in figure \eqref{fideli} we plot the fidelity of
each state with the vacuum state (to be asymptotically reached) as
a function of time. The fidelity $F(\rho_A,\lv 0\ra\la 0 \rv) =
\sqrt{\la 0\rv\rho_A\lv 0\ra}$ is given by \be F^2(\rho_A,\lv
0\ra\la 0 \rv) = \frac{2 \ei^{-|\beta(t)|^2}}{N^2} \Big{(} 1+
f(\beta_0,t) \cos \theta \Big{)}, \ee where $f(\beta_0,t)$ is
given by \eqref{ft}, $N $ is the normalization
\eqref{normalizacao} and $\beta(t)$ is given by \eqref{betat}.

Initially the odd superposition state does not contain the vacuum
in its structure. However, since this state is a fixed point of
the dynamics (its pointer state), it needs to be populated. As can
be gathered from the figure, the vacuum state is rapidly
populated. As for the even superposition state, if $\beta_0
\lesssim 1.14$ (for our choice of parameters), the initial
probability of finding the vacuum in that state is always larger
than 50\%. There comes the difference with the odd superposition
state. While the latter rapidly populates the vacuum state, the
even superposition state starts by populating other states
maintaining the vacuum population practically unchanged. After
these initial different transient times both states have similar
dynamics, evolving both to the asymptotic state $\lv 0\ra \la
0\rv$.

Next, we show that the characteristic time of squeezing effects is
shorter than that for interference effects. The tool we use here to
investigate squeezing in the quadratures is the determinant of the
covariance matrix \begin{align} \label{det}
D(t) &= \textrm{Det}
\begin{pmatrix}
   \sigma_{pp} & \sigma_{qp}\\
   \sigma_{qp} & \sigma_{qq}
\end{pmatrix},
\end{align} where $\sigma_{qq} = \la x^2 \ra- \la x\ra^2$,
$\sigma_{pp} = \la p^2 \ra- \la p\ra^2$ and
$\sigma_{qp} . \sigma_{pq} = \frac{1}{4}\la xp + px\ra^2$. For CSS, we have
\bq \lan x^2 \ran
&=& \frac{1}{2 \omega} \Bigg{\{} 1+ \beta^2 + \beta^{* 2}  \nonumber
\\&& + \frac{4 |\beta|^2}{N^2} \left[ 1-\cos(\theta) \ei^{-2|\beta_0|^2}\right]
\Bigg{\}} \nonumber \\ \lan p^2 \ran &=& -\frac{\omega}{2}
\Bigg{\{} -1+ \beta^2 + \beta^{* 2}
\nonumber \\
&&  - \frac{4 |\beta|^2}{N^2}\left[ 1-\cos(\theta)
\ei^{-2|\beta_0|^2}\right] \Bigg{\}} \nonumber \\ \lan xp+px \ran &=& -i(\beta^2-\beta^{* 2}). \nonumber \eq since
the non-diagonal terms precisely cancel the rapid oscillations due
to the free field frequency. Note that, for this case, $\la x^2
\ra = \sigma_{qq}$ and $\la p^2 \ra = \sigma_{pp}$

As can be noted from the expressions for $\la x^2 \ra$ and $\la
p^2 \ra$ (showed explicitly in \cite{knightsqueeze}) the odd
superposition state will never exhibit squeezing. However, the
even superposition is always squeezed.

\begin{figure} [!h]
\begin{center}
\includegraphics [bb=0 45 300 286, scale=0.6] {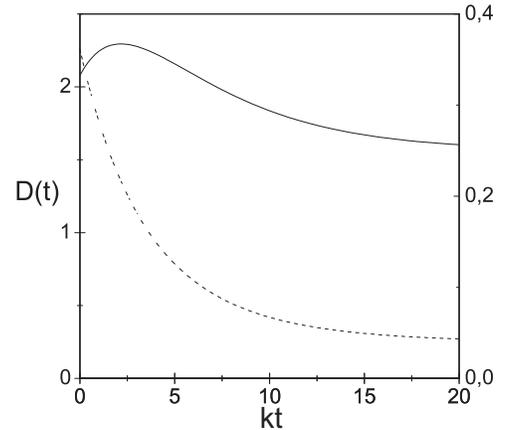}
\end{center}
\caption{Time evolution of the determinant of the covariance
matrix for the even superposition state (solid line, right scale)
and for the odd superposition state (dashed line, left scale).
Parameters: $\beta_0 = 0.8$, $\omega = 1$, $k = 0.1$.}
\label{figura5}
\end{figure}

\begin{figure} [!h]
\begin{center}
\includegraphics [bb=0 45 300 286, scale=0.6] {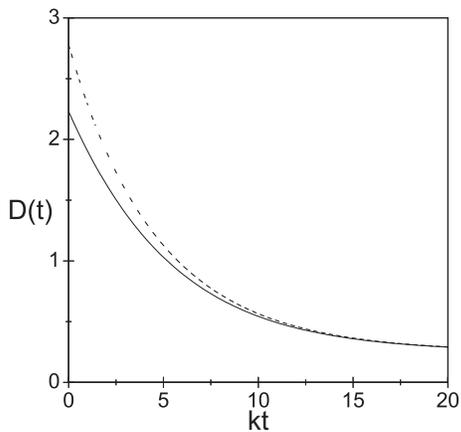}
\end{center}
\caption{Time evolution of the determinant of the covariance
matrix for the even superposition state (solid line) and for the
odd superposition state (dashed line). Parameters: $\beta_0 =
1.5$, $\omega = 1$, $k = 0.1$.} \label{figura6}
\end{figure}

In figures \eqref{figura5} and \eqref{figura6} we show the
determinant of the covariance matrix for the even and odd
superposition states. As can be seen from the figure, the
squeezing of the even superposition state (``filtered'' by the
determinant) increases, reaching a maximum value and then
following the dissipative dynamics which will take it to the
vacuum state. The time of the maximum value of the determinant
(and of the squeezing ``visibility'') depends on $\beta_0$ as
follows \be \label{tccat} t_{c}^{S} = - \frac{1}{2k}
\ln\left[\frac{\sinh\left(2|\beta_{0}|^{2}\right)}
{4|\beta_{0}|^{2}\cos \theta}\right], \ee and the effect is only
visible for small enough values of $\beta_0$, i. e., only if \be
\label{catcondition} 0 < \frac{\sinh\left(2|\beta_{0}|^{2}\right)}
{4|\beta_0|^2\cos \theta} < 1. \ee For the even states
\be 0 \leq t_{c}^{S} \leq \tau ,\ee where $\tau$ is the
decoherence time (note that for the odd CSS the characteristic time does not have physical interpretation, it acquires imaginary values). For large values of $\beta_0$ the squeezed
quadrature is essentially constant, and the determinant of the
covariance matrix is very similar for both the even and odd
superposition states. We remark that the visibility of the effect
is a consequence of two factors: the initial conditions must
obey the above inequality and the characteristic times must be
experimentally ``available''.

\subsection{Oscillating photon distribution} \label{catosci}

\begin{figure} [!h]
\begin{center}
\includegraphics [scale=0.45] {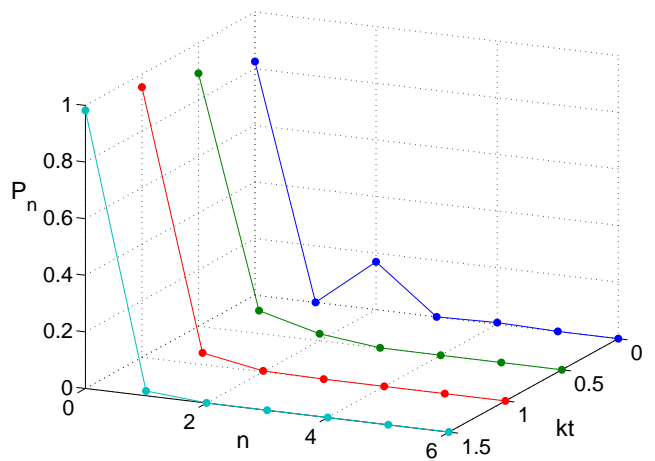}
\end{center}
\caption{Time evolution of the photon number distribution for an
even superposition state ($\theta = 0$). Parameters: $\beta_0 =
0.8$, $\omega = 1$, $k = 0.1$.} \label{figure8}
\end{figure}

\begin{figure} [!h]
\begin{center}
\includegraphics [scale=0.45] {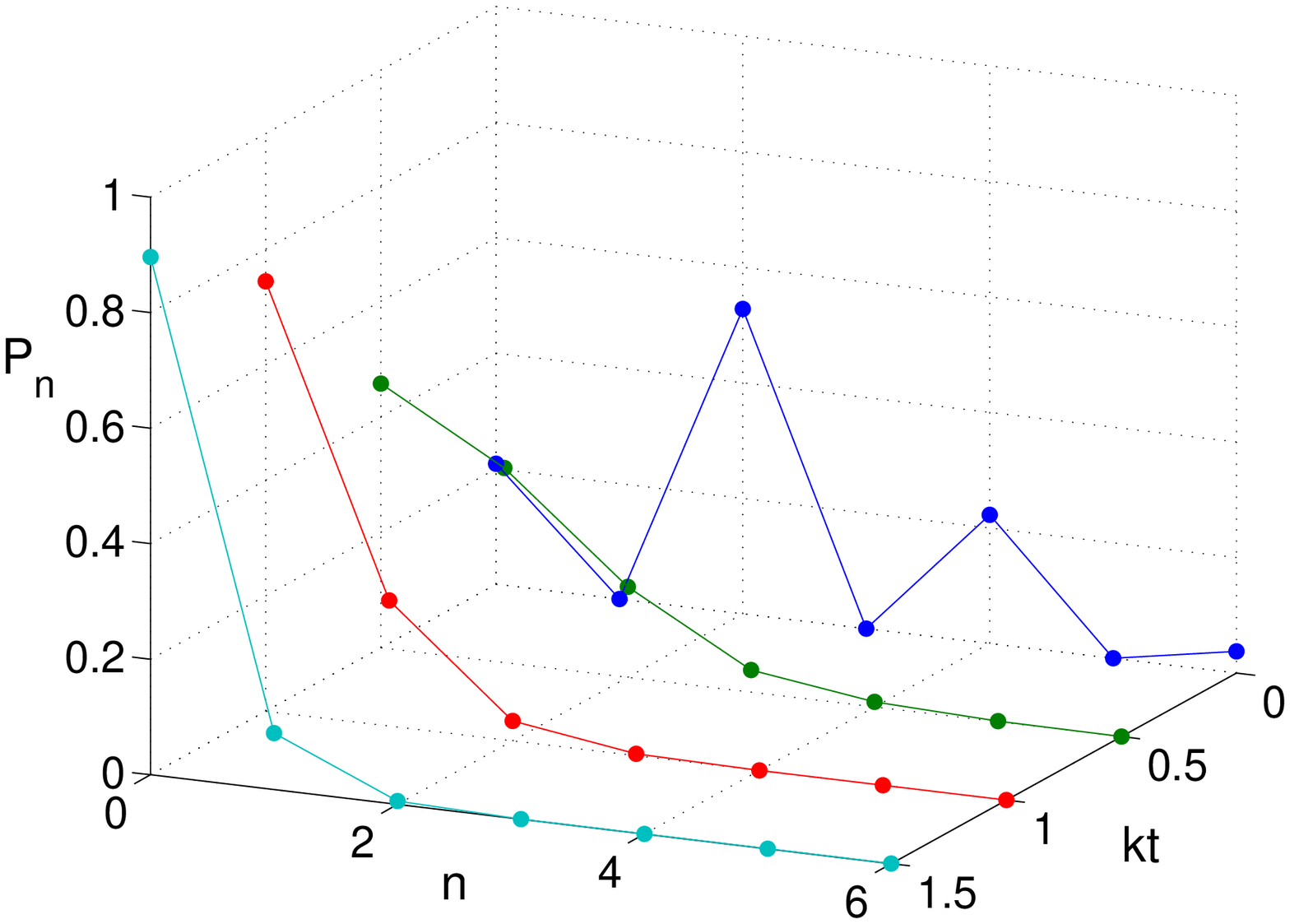}
\end{center}
\caption{Time evolution of the photon number distribution for an
even superposition state ($\theta = 0$). Parameters: $\beta_0 =
1.5$, $\omega = 1$, $k = 0.1$.} \label{figure9}
\end{figure}

\begin{figure} [!h]
\begin{center}
\includegraphics [scale=0.45] {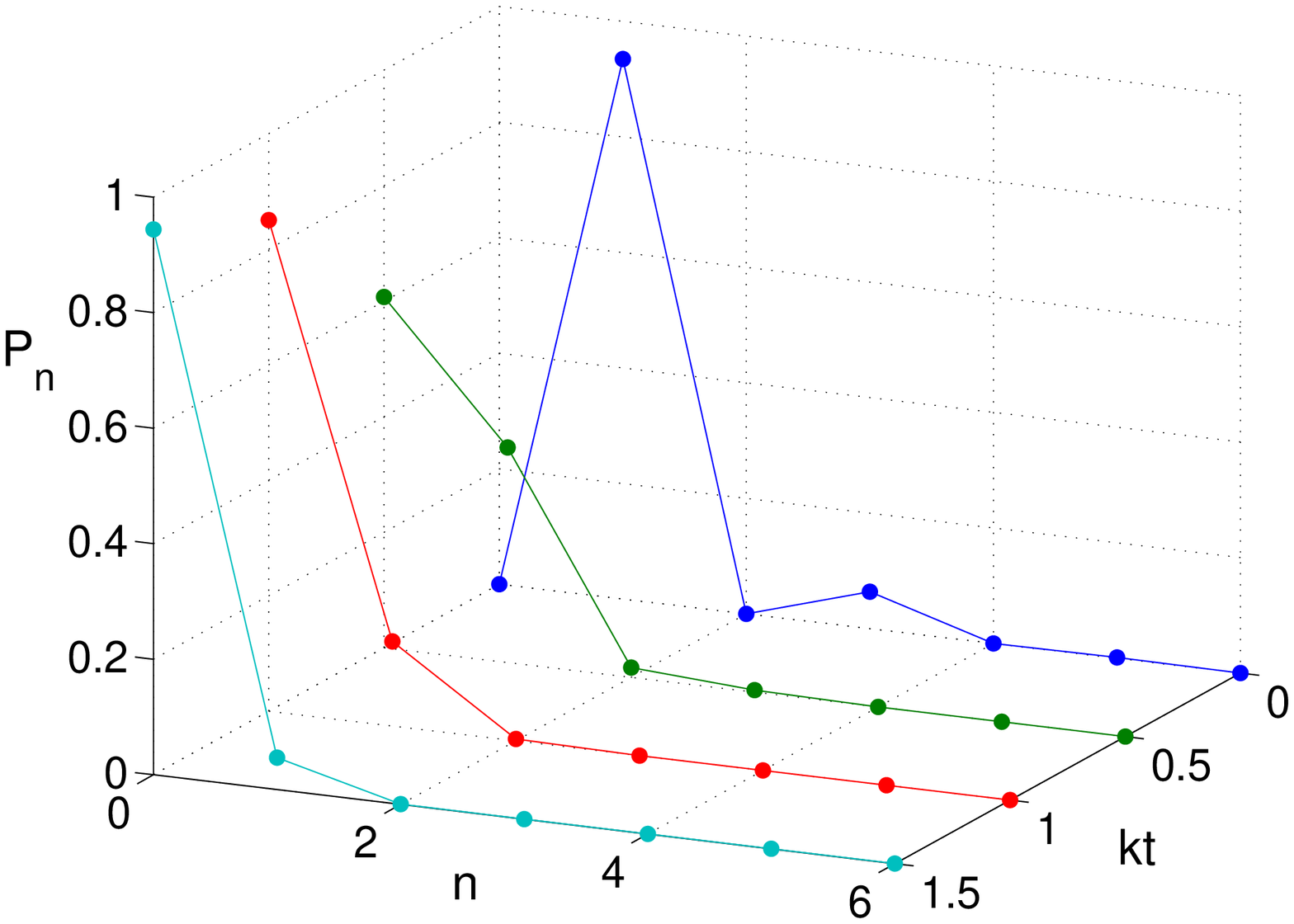}
\end{center}
\caption{Time evolution of the photon number distribution for an
odd superposition state ($\theta = \pi$). Parameters: $\beta_0 =
0.8$, $\omega = 1$, $k = 0.1$.} \label{figure10}
\end{figure}

\begin{figure} [!h]
\begin{center}
\includegraphics [scale=0.45] {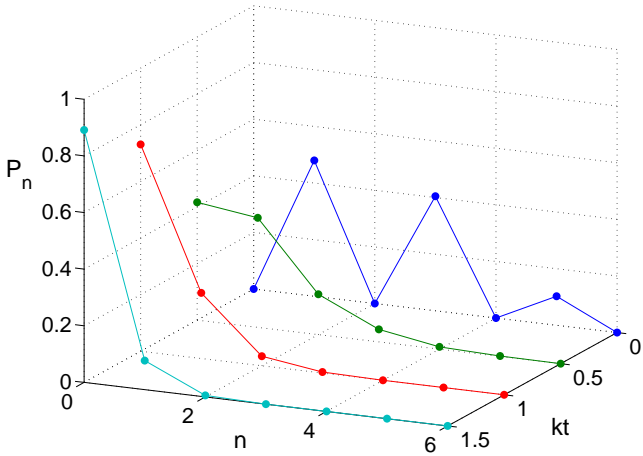}
\end{center}
\caption{Time evolution of the photon number distribution for an
odd superposition state ($\theta = \pi$). Parameters: $\beta_0 =
1.5$, $\omega = 1$, $k = 0.1$.} \label{figure11}
\end{figure}

The time evolution of the photon distribution for the even and odd
superposition states for different values of $\beta_0$ is depicted
in figures \eqref{figure8}-\eqref{figure11}. The analytic
expression for these curves is given by ($P_n = \lan n | \rho | n \ran$)\be P_n =
\ei^{-|\beta(t)|^2} \frac{2 |\beta(t)|^{2 n}}{N^2 n!} \left[ 1+
(-1)^n f(\beta_0, t) \cos\theta \right]. \ee Note that the
dissipative dynamics will tend to destroy the initial parity of
the states. The characteristic time is approximately the same as
that for the squeezing of the even superposition state. The solid
line is intended to guide the eye. Also, it is well known that while
coherent states have Poissonian photon distribution, even (odd)
superposition states have super(sub)-Poissonian distributions.
This can be measured by the Mandel parameter $Q$ defined as \be Q
= \frac{\la (\Delta n)^2 \ra - \la n \ra}{\la n \ra}. \ee Here,
$\la n \ra$ and $\la (\Delta n)^2\ra$ are the average and the
variance of photon number in the field state, respectively. If the
distribution is sub-Poissonian, i. e. $Q < 0$, the state state
\emph{is}~ a quantum one. If $Q \geq 0$ however, no definite
statement can be made. For example, in our case, the even CSS is ``as
quantum'' as the odd one, despite presenting super-Poissonian
statistics (just like ``classical'' light) \cite{knightsqueeze}, which evidences that in order to decide
whether a state is quantum or not, most likely more than one pertinent observables
should be measured. The exception to this case is the set of states that present negative Wigner
function in which case a simple measurement of this negativity is enough to preclude any classical
analog.

\subsection{von Neumann entropy} \label{catent}

\begin{figure} [!h]
\begin{center}
\includegraphics [bb=40 40 300 286, scale=0.6] {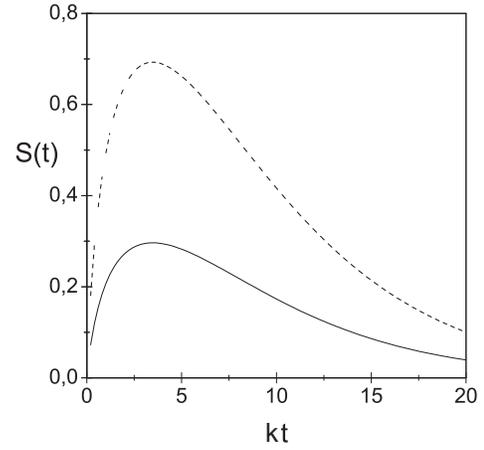}
\end{center}
\caption{von Neumann's entropy for the even superposition state
($\theta = 0$, solid line) and for the odd superposition state
($\theta = \pi$, dashed line). Parameters: $\beta_0 = 0.8$,
$\omega = 1$, $k = 0.1$.} \label{figura12}
\end{figure}

\begin{figure} [!h]
\begin{center}
\includegraphics [bb=40 40 300 286, scale=0.6] {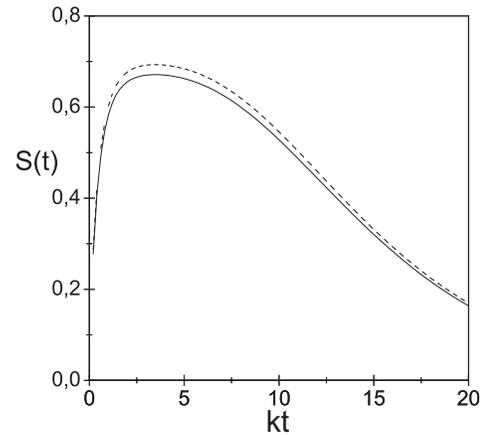}
\end{center}
\caption{von Neumann's entropy for the even superposition state
($\theta = 0$, solid line) and for the odd superposition state
($\theta = \pi$, dashed line). Parameters: $\beta_0 = 1.5$,
$\omega = 1$, $k = 0.1$.} \label{figura13}
\end{figure}

Having obtained the eigenvalues of the density matrices of even
and odd superposition states, it is a simple matter to calculate
von Neumann's entropy, which is depicted in figures
\eqref{figura12}-\eqref{figura13} for different values of
$\beta_0$ . The von Neumann entropy is given by \be S[\rho] = -
\tr (\rho \ln \rho) \ee and for the superposition states
considered \be S[\rho_A] = - p_o \ln p_o - p_e \ln p_e .\ee Note
that the entropy increases up to the decoherence time,
when it starts decreasing back to zero, which is the entropy of the
asymptotic state of the dissipative reservoir, the vacuum. The
shape of the curve changes for large values of $\beta_0$ and the
decoherence time is smaller, as expected.

\section{Gaussian states: displaced, squeezed, thermal
states (GSS)} \label{DSTS}

We start with the Liouvillian given in Eq. \eqref{liouvilliano1}, but now we
we extend the calculations in order to include temperature, i.e., we
consider $\bar{n}_B \ge 0$.

If the initial state is a Gaussian, it can be written as \be
\label{gaussini} \rho_B = \mathcal{D}(\alpha_0)
\mathcal{S}(r_0,\phi_0) \rho_{\nu_0}
\mathcal{S}^\dagger(r_0, \phi_0) \mathcal{D}^\dagger (\alpha_0)
\ee where $\mathcal{D}(\alpha)$ is the displacement operator,
$\mathcal{S}(r,\phi)$ is the squeezing operator, the subscript 0
denotes initial values and $\rho_\nu$ is the thermal density
operator with average number of excitations $\nu$. More explicitly
we have \bq \mathcal{D}(\alpha) &=& \exp(\alpha a^\dagger -
\alpha^* a)
\\ \mathcal{S}(r,\phi) &=& \exp \Big{(} \frac{1}{2} r \ei^{i \phi}
a^{\dagger 2} - \frac{1}{2} r \ei^{-i \phi} a^2 \Big{)}
\\ \rho_\nu &=& \frac{1}{1+\nu} \exp \left[ \ln \left(
\frac{\nu}{\nu+1} \right) a^\dagger a \right] \eq

We remark that $\nu$ is \emph{not} the average number of
excitations, but rather the average number of \emph{thermal}
excitations. The evolution preserves the Gaussian character of
the density operator, and the parameters acquire a time
dependence, i. e., the state becomes \be \label{gaussiana}
\rho_B = \mathcal{D}(\alpha) \mathcal{S}(r,\phi)
\rho_{\nu} \mathcal{S}^\dagger(r, \phi) \mathcal{D}^\dagger
(\alpha) \ee where \bq
\label{alpha1} \alpha (t) &=& \alpha_0 \ei^{-(i \omega + k) t} \\
\phi(t) &=& \phi_0 - 2 \omega t \\ \nu(t) &=& \sqrt{x^2(t) -
\left[(\nu_0+\frac{1}{2})
\sinh (2 r_0) \ei^{-2 k t} \right]^2}-\frac{1}{2} \nonumber \\
\\x(t) &=& \left( \nu_0 + \frac{1}{2}\right) \cosh(2 r_0) \ei^{-2 k
t} + \left( \overline{n}_B + \frac{1}{2} \right) (1-\ei^{-2 k t})
\nonumber \\ \\ r(t)&=& \frac{1}{4} \ln \left[ \frac{(\nu_0 +
\frac{1}{2}) \ei^{2 r_0}+ (\overline{n}_B + \frac{1}{2}) (\ei^{2 k
t} -1) }{(\nu_0 + \frac{1}{2}) \ei^{-2 r_0}+ (\overline{n}_B +
\frac{1}{2}) (\ei^{2 k t} -1) } \right] \nonumber  \label{rt}
\\\eq

The Robertson-Schroedinger determinant (or the covariance matrix)
and the von Neumann entropy are related by \bq \label{detdsts}
D(t) &=& \left[ \nu(t) +\frac{1}{2} \right]^2 \\S[\rho(t)] &=&
\left[\nu(t)+1\right] \ln \left[\nu(t)+1\right] - \nu(t) \ln
\nu(t). \nonumber
\\ \label{entdsts} \eq

Note that, in the case of Gaussian states, the entropy is
completely determined by a relationship between quadratures, given
by $D(t)$, and is always analytical. Note also that, differently
from the superposition states, the entropy is independent of the
optical field intensity, which may turn on an experimental
advantage. Here, optical field intensity means the displacement.

\subsection{Squeezing} \label{gausssquee}

The Wigner function of the GSS is always positive. The evolved
Wigner function for a GSS described by  the Liouvillian
\eqref{liouvilliano1} acting on the Gaussian initial state
\eqref{gaussini} is \bq W(q,p) &=& \sum_l \frac{1}{\pi}
\frac{\left(-|F_3|\nu\right)^l}{(\nu+1)^{l+1}}   \nonumber \\ &&
\times L_l \Big{[} 2 \Big{(} \frac{(x-x_0)^2}{F_4^2} + \frac{(F_4
F_5)^2}{4} \Big{)}\Big{]}  \\ && \label{wignergauss} \times
\frac{F_4}{|F_1|} \exp\left[-\frac{(x- x_0)^2}{F_4^2} -\frac{(F_4
F_5)^2}{4}\right] , \nonumber \eq where $L_l(x)$ is the Laguerre
function of $l$ order and argument $x$ and we define \cite{nieto}
\bq F_1 &=& \cosh r + \ei^{i \phi} \sinh r, \nonumber
\\ F_2 &=& \frac{1- i \sin \phi \sinh r (\cosh r + \ei^{i \phi}
\sinh r)}{(\cosh r + \cos \phi \sinh r)(\cosh r + \ei^{i \phi}
\sinh r)}, \nonumber \\ F_3 &=& \frac{\cosh r + \ei^{-i \phi} \sin
\phi \sinh r}{\cosh r + \ei^{i \phi} \sin \phi \sinh r}, \nonumber \\
F_4 &=&\sqrt{\cosh^2 r + \sinh^2 r + 2 \cos \phi \cosh r \sinh r},
\nonumber \\ F_5 &=& 2 (p+p_0)- i (x-x_0)(F_{2}^{*}-F_2). \eq The
time evolution of the parameters are given by Eq.
\eqref{alpha1}-\eqref{rt}. The Wigner function above can also be
written  in a closed Gaussian form (since the dynamics does not
change its Gaussian character): \bq W(x,p) &=& \frac{1}{\pi (\nu +
\frac{1}{2})} \exp \Big{\{} \frac{\sin \phi \sinh(2 r)}{\nu +
\frac{1}{2}} x p - \nonumber \\ && - \frac{\cosh(2 r)}{2 \nu +1}
(1-\tanh(2 r) \cos \phi) x^2 \nonumber \\ && - \frac{\cosh(2 r)}{2
\nu +1} (1+\tanh(2 r) \cos \phi) p^2
 \Big{\}}. \eq

\begin{figure} [!h]
\begin{center}
\includegraphics [bb=40 40 300 286, scale=0.6]{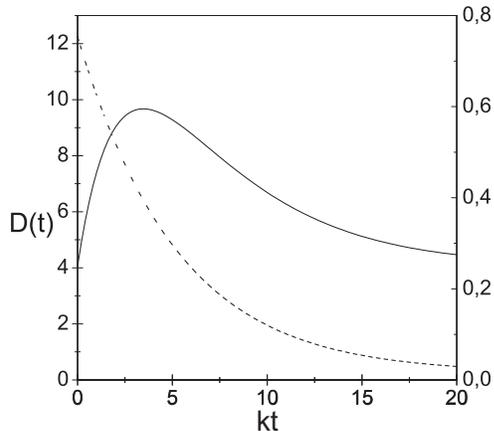}
\end{center}
\caption{Time evolution of the determinant of the covariance
matrix for the GSS. Parameters: $\omega = 1$, $n_B = 0$, $k =
0.1$, $r_0 = 1$, $\nu_0 = 0$ (solid line, right scale) and $\nu_0
= 3$ (dashed line, left scale).} \label{detgaussfig}
\end{figure}

We can access the squeezing by looking at the determinant of the
covariance matrix \eqref{detdsts} (and consequently looking at the
entropy \eqref{entdsts}). In figure \eqref{detgaussfig} we show
the determinant of the covariance matrix versus time. Its behavior
(and the discussion) is very similar to the one of the even
superposition state showed before. Here, the time of the maximum
value of the determinant is (this result was also found in ref. \cite{marian2}, for the linear entropy) \bq \label{tcgauss} t_{c}^{G} &=& (2
k)^{-1} \Big{\{} \ln 2 -
\ln \Big{[} \frac{2 \bar{n}_B +1}{d}   \\
&\times&(2 \nu_0 \cosh (2 r_0) + \cosh (2 r_0) -2 \bar{n}_B -1)
\Big{]} \Big{\}}\nonumber\eq where \bq d &=&
2\cosh\left(2r_0\right)
\left[\bar{n}_B\left(\nu_0+1\right) +\nu_0\left(\bar{n}_B + 1\right)+\frac{1}{2}\right] \nonumber \\
&&-2\left(\bar{n}_B+\frac{1}{2}\right)^2-2\left(\nu_0+\frac{1}{2}\right)^2.
\nonumber \eq We remark that we assume the characteristic time to
be positive, i. e. if $t_c^G > 0 \longrightarrow t_c^G \in
\mathbb{R}$. The ``quantum properties'' present in the state are
visible for $\nu_0$ satisfying \cite{caroleeu} \be \label{inequal}
\nu_0 < \frac{1}{2} \left[2 \bar{n}_B \cosh\left(2
r_0\right)+\cosh\left(2 r_0\right)-1 \right] .\ee Note that, given
an initial condition $\nu_0$, the temperature of the reservoir
cannot be very high, therefore the system will tends to
equilibrium rapidly. As a matter of fact, {\it given an initial
condition $\nu_0$}, the temperature must satisfy \be \bar{n}_B <
\frac{1}{2} \left[2 \nu_0 \cosh\left(2 r_0\right)+\cosh\left(2
r_0\right)-1 \right],\ee such that an increasing in $D(t)$ (or in
the entropy) be visible. Of course, this visibility is also a
consequence of two factors: the initial conditions need to obey
the above inequality \eqref{inequal} {\it and} the time scale
\eqref{tcgauss} needs to be experimentally accessible.

Equation \eqref{inequal} establishes an upper bound on the initial
``impurity'' of the state such that, even under a dissipative
environment, squeezing can be accessed.

It is rather remarkable that this characteristic time is
independent of the field intensity. This has been shown
numerically in reference \cite{numerico}. For our choice of
parameters, $t_{c}^{G}$ is comparable to $t_{c}^{S}$
(superposition time scale). However, it is possible to obtain
$t_{c}^{G} \gg t_{c}^{S}$, \textit{e.g.}, by simply increasing the
field intensity of the coherent superposition state. As an
example, if we impose $\lan a^\dagger a \ran_{GSS} = \lan
a^\dagger a \ran_{CSS}$ and choose the following parameters:
$\omega = 1$, $n_B = 0$, $k =0.1$, $\nu_0 = 0$, $|\beta_0| = 0.8$,
$\alpha_0 = 0.0$ and $\theta = 0$, the squeezing factor must be
$r_0 \simeq 0.73$ and the characteristic times will respect \bq  0
< t_c^S < t_c^G < \tau, \eq where $\tau$ is the decoherence time
for the CSS, $t_c^S$ and $t_c^G$ is given by \eqref{tccat} and
\eqref{tcgauss} respectively.

\subsection{Oscillating photon distribution} \label{gaussosci}

Analyzing the displaced squeezed Gaussian states, one can see that
they also have super-Poissonian statistic (as the even
superposition state). This is measured by the Mandel parameter:
for a GSS with the dynamics given by \eqref{liouvilliano1} we
always have $Q \geq 0$. For a GSS it is known that the photon
distribution is (given in \cite{marian2,marian}) \bq P_n &=& \pi
Q(0) (-1)^n 2^{-2 n} (\tilde{A}+|\tilde{B}|)^n \nonumber \\ &&
\times \sum_{k=0}^{n} \frac{1}{k! (n-k)!}\Big{[} \frac{\tilde{A}-
|\tilde{B}|}{\tilde{A}+|\tilde{B}|} \Big{]}^k \nonumber
\\ && \times H_{2 k} \Bigg{[} i \frac{\Im (\tilde{C} \ei^{-i
\frac{\phi}{2}})}{\sqrt{\tilde{A}-|\tilde{B}|}} \Bigg{]} \nonumber \\
&& \times H_{2 n - 2 k} \Bigg{[} i \frac{\Re (\tilde{C} \ei^{-i
\frac{\phi}{2}})}{\sqrt{\tilde{A}+|\tilde{B}|}} \Bigg{]} \eq where
$H_j$ is the $j$-order Hermite polynomial and \bq \pi Q(0) &=&
[(1+A)^2-|B|^2]^{1/2} \nonumber \\ && \times \exp \Bigg{\{} -
\frac{(1+A) |C|^2 + \frac{1}{2} [B(C^*)^2+B^* C^2]}{(1+A)^2-|B|^2} \Bigg{\}} \nonumber \\
\eq  where \bq A &=& \nu + (2 \nu +1) \sinh^2 r \\ B &=& -(2 \nu
+1) \ei^{i \phi} \sinh r \cosh r \\ C &=& \alpha \eq and finally
\bq \tilde{A} &=& \frac{\nu (\nu + 1)}{\nu^2 + (\nu +
\frac{1}{2})[1+\cosh (2 r)]} \\ \tilde{B} &=& - \frac{\ei^{i \phi}
(\nu + \frac{1}{2}) \sinh (2 r)}{\nu^2 + (\nu +
\frac{1}{2})[1+\cosh (2 r)]} \\ \tilde{C} &=& \frac{C [\frac{1}{2}
+ (\nu + 1/2) \cosh (2 r)]- C^* \ei^{i \phi} (\nu + \frac{1}{2})
\sinh (2 r)}{\nu^2 + (\nu + \frac{1}{2})[1+ \cosh (2 r)]}.
\nonumber \\ \eq

\begin{figure} [!h]
\begin{center} 
\includegraphics [scale=0.45] {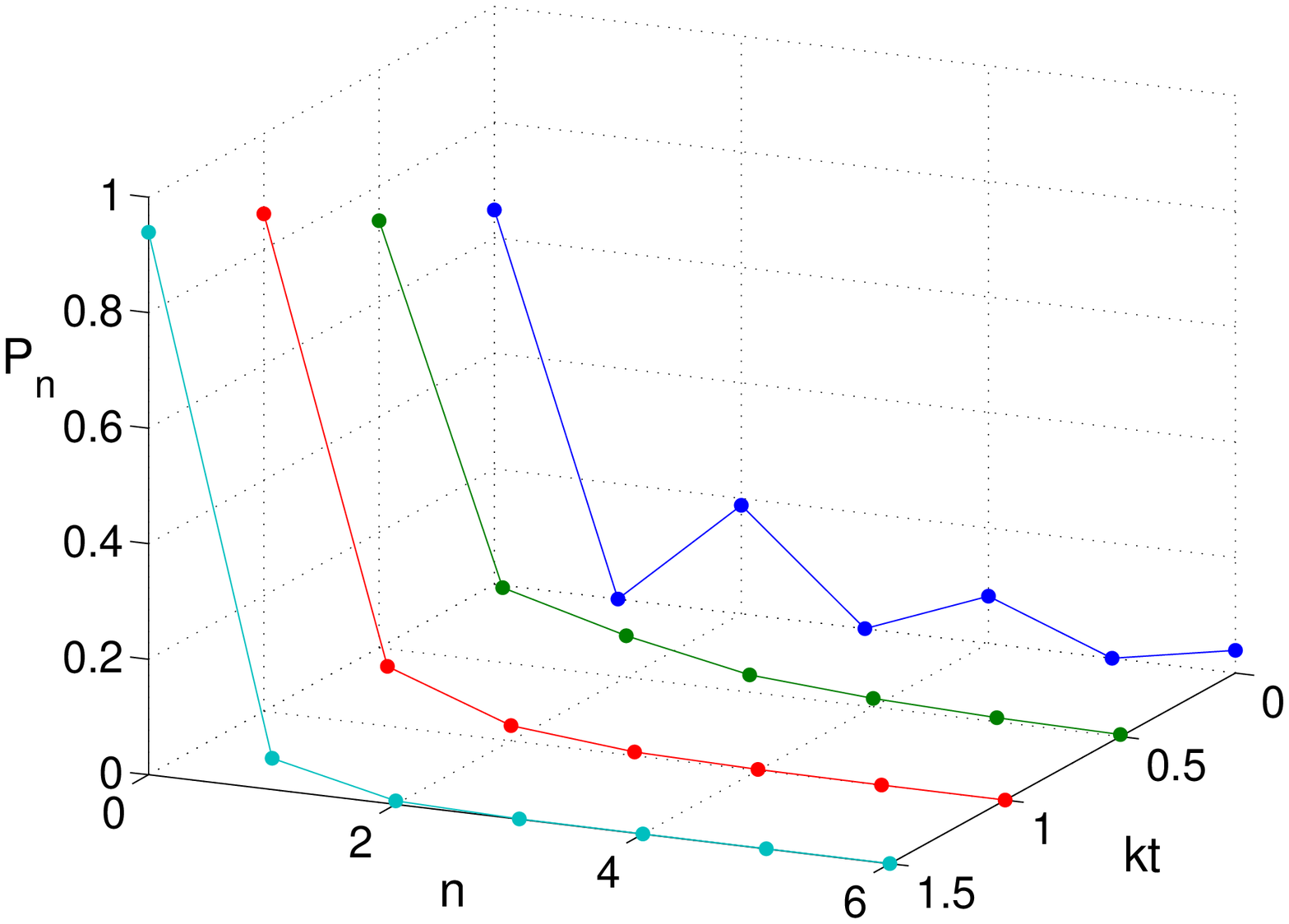}
\end{center}
\caption{Photon number distribution for a GSS with $\alpha_0=0$,
$\phi_0=0$, $r_0 = 1$, $k = 0.1$, $\nu_0=0$ and $\bar{n}_B=0$.}
\label{pngaussp}
\end{figure}

\begin{figure} [!h]
\begin{center}
\includegraphics [scale=0.45] {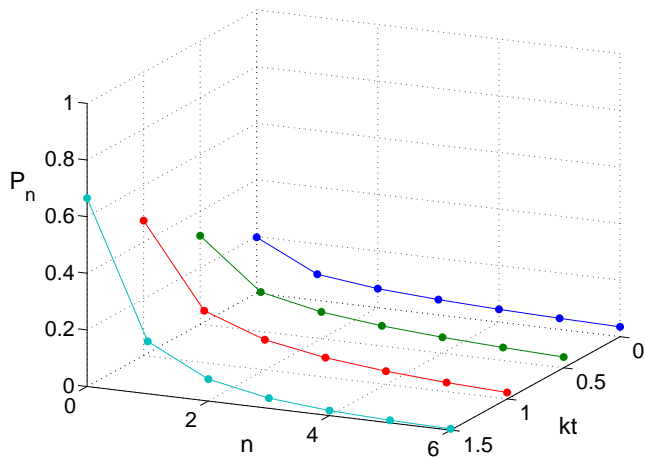}
\end{center}
\caption{Photon number distribution for a GSS with $\alpha_0=0$,
$\phi_0=0$, $r_0 = 1$, $k = 0.1$, $\nu_0=3$ and $\bar{n}_B=0$.}
\label{pngaussg}
\end{figure}

The time evolution of $P_n$ is presented in figures
\eqref{pngaussp}-\eqref{pngaussg}. It is clear that, if the states
respect \eqref{inequal} oscillations in the photon distribution
are observable, else the photon distribution look like a
``thermal'' one.

Following the idea of \cite{polones} the second order correlation
function for Gaussian Squeezed States (GSS) is a ``quantum
witness'' \be g^{(2)} (0) =\frac{\la a^\dagger a^\dagger aa
\ra}{\la a^\dagger a \ra^2}. \ee If $g^{(2)} > 3$ the GSS {\it is}
quantum, if $g^{(2)} \leq 3$ the state is classical. We see that
for values respecting \eqref{inequal} the state is quantum
according this criterion. The Mandel parameter $Q$ for a GSS
shows, like the even superposition state, super-Poissonian
statistic $Q>0$, and when $t \rightarrow \infty$ the statistic
tends to Poissonian.

\subsection{von Neumann entropy} \label{gaussent}

\begin{figure} [!h]
\begin{center} 
\includegraphics [bb=40 40 300 300, scale=0.6] {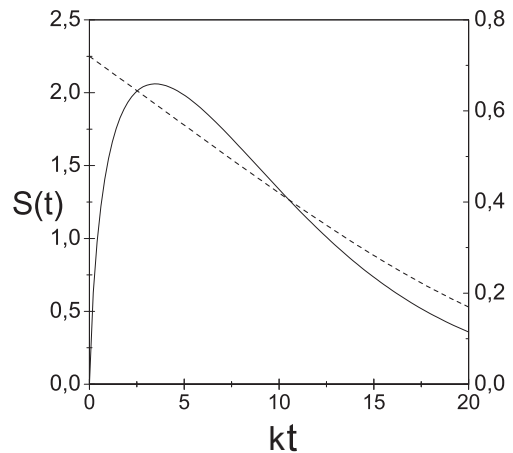}
\end{center}
\caption{Time evolution of the von Neumann entropy for the GSS.
Parameters: $\omega = 1$, $n_B = 0$, $k = 0.1$, $r_0 = 1$, $\nu_0
= 0$ (solid line, right scale) and $\nu_0 = 3$ (dashed line, left
scale).} \label{entgauss}
\end{figure}

In a recent work \cite{caroleeu} we studied some characteristics
of the GSS, including photon number distribution, Wigner function
and von Neumann entropy. We show the results here for the purpose
of comparison with the superposition states (similar behaviour was found also in \cite{marian3}, where they studied the 2-entropy of a single-mode field initially in a number state). In figure
\eqref{entgauss} we show the result for the von Neumann entropy
for the GSS, for different values of $\nu_0$. The characteristic
time for the entropy is the same, in this case, as the
characteristic time for squeezing, and we can conclude that the
same condition (i. e., equation \eqref{inequal}) holds here too.
Note also that the entropy (and any other observable that depends
only on $\nu(t)$) is {\it independent} of the field strength
$\alpha(t)$ and the squeezing phase $\phi(t)$.

\section{GSS versus superposition states -- Fidelity}
\label{secfid}

In the previous sections we briefly review some properties of
coherent superposition states and of GSS, and how these properties
evolve in time. We compare these states by graphically analyzing
the evolution of those properties and, although qualitatively, we
could get some conclusions and made some comparisons between the
studied states.

Now we present a more quantitative comparison between the states,
through the fidelity, defined by \be F = \tr
\sqrt{\sqrt{\rho_A}\rho_B\sqrt{\rho_A}} \ee where $\rho_A$ and
$\rho_B$ are \eqref{rhobeta} and \eqref{gaussiana} respectively.
With this result we can quantify how the states ``look like'' each
other.

The expression for the fidelity is \be F=
\sqrt{\lambda_+}+\sqrt{\lambda_-},\ee
where \be\lambda_\pm=\frac{b + c \pm \sqrt{(b-c)^2+4|d|^2}}{2}\ee
and \bq b &=& p_e \lan e |\rho_B | e \ran \nonumber
\\ &=& \frac{p_e}{N_{e}^{2}} \Big{\{} \lan \beta |\rho_B| \beta
\ran + \lan - \beta |\rho_B| - \beta \ran + 2 \Re \lan - \beta |\rho_B| \beta \ran \Big{\}} \nonumber \\
\\  c &=& p_o \lan o | \rho_B | o \ran \nonumber \\ &=& \frac{p_o}{N_{o}^{2}} \Big{\{} \lan \beta |\rho_B| \beta
\ran + \lan - \beta |\rho_B| - \beta \ran - 2 \Re \lan - \beta |\rho_B| \beta \ran \Big{\}} \nonumber \\
\\ d &=& \sqrt{p_e p_o} \lan e |\rho_B | o \ran \nonumber \\ &=&
 \frac{\sqrt{p_e p_o}}{N_{e} N_o} \Big{\{} \lan \beta |\rho_B| \beta
\ran - \lan - \beta |\rho_B| - \beta \ran + 2 i \Im \lan - \beta
|\rho_B| \beta \ran \Big{\}}. \nonumber \\ \eq

The other terms are \bq \lan \beta | \rho_B | \beta \ran &=&
\frac{1}{\sqrt{(\nu+1)^2 \cosh^2 r - \nu^2 \tanh^2 r}} \nonumber
\\ && \times \exp \Big{\{} \frac{(2 \nu + 1) \tanh r~ \Re [\eta^2]}
{(\nu + 1)^2 - \nu^2 \tanh^2 r} \Big{\}} \nonumber \\ && \times
\exp \Big{\{}- \frac{|\eta|^2 [(\nu + 1) + \nu \tanh^2 r]}{(\nu +
1)^2 - \nu^2 \tanh^2 r} \Big{\}} \nonumber \\ \eq \bq \lan - \beta
| \rho_B | - \beta \ran &=& \frac{1}{\sqrt{(\nu+1)^2 \cosh^2 r -
\nu^2 \tanh^2 r}} \nonumber
\\ && \times \exp \Big{\{} \frac{(2 \nu + 1) \tanh r~ \Re [\zeta^2]}
{(\nu + 1)^2 - \nu^2 \tanh^2 r}
\Big{\}} \nonumber
\\ && \times \exp \Big{\{}- \frac{|\zeta|^2 [(\nu + 1) +
\nu \tanh^2 r]}{(\nu + 1)^2 - \nu^2 \tanh^2 r} \Big{\}} \nonumber
\\ && \eq \bq \lan - \beta | \rho_B | \beta \ran
&=& \frac{1}{\sqrt{(\nu+1)^2 \cosh^2 r - \nu^2 \tanh^2 r}}
\nonumber \\ && \times \exp \Big{\{} \frac{(2 \nu + 1) \tanh r~
[\zeta^2 + \eta^2]} {2 (\nu + 1)^2 - 2 \nu^2 \tanh^2 r} \Big{\}}
\nonumber
\\ && \times \exp \Big{\{}- \frac{|\zeta|^2 + |\eta|^2}{2}
\nonumber \\ && -  \frac{\zeta \eta \nu (\nu + 1)} {(\nu + 1)^2
\cosh^2 r - \nu^2 \sinh^2 r} \Big{\}} \nonumber
\\ && \eq where we define \bq \zeta = (\beta^* + \alpha^*) \ei^{i \frac{\phi}{2}}
\\ \eta = (\beta - \alpha) \ei^{-i \frac{\phi}{2}}. \eq

\begin{figure} [!h]
\begin{center}
\includegraphics [scale=0.6] {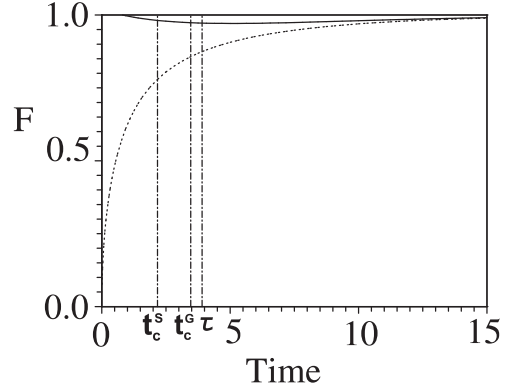}
\end{center}
\caption{Fidelity between the coherent superposition states and
the GSS. The parameters used: $\beta_0 = 0.8$, $\omega=1$,
$k=0.1$, $r_0 =1$, $\alpha_0 \simeq 0.0$, $\phi_0 = 0$, $n_B = 0$,
$\nu_0 = 0$, $\theta = 0$ (solid line) and $\theta = \pi$ (dashed
line). The initial value of F for the even CSS is greater than 1
due to limitations in the numeric routine used.}
\label{fidelidade1}
\end{figure}

\begin{figure} [!h]
\begin{center}
\includegraphics [bb=39 53 345 293 ,scale=0.6] {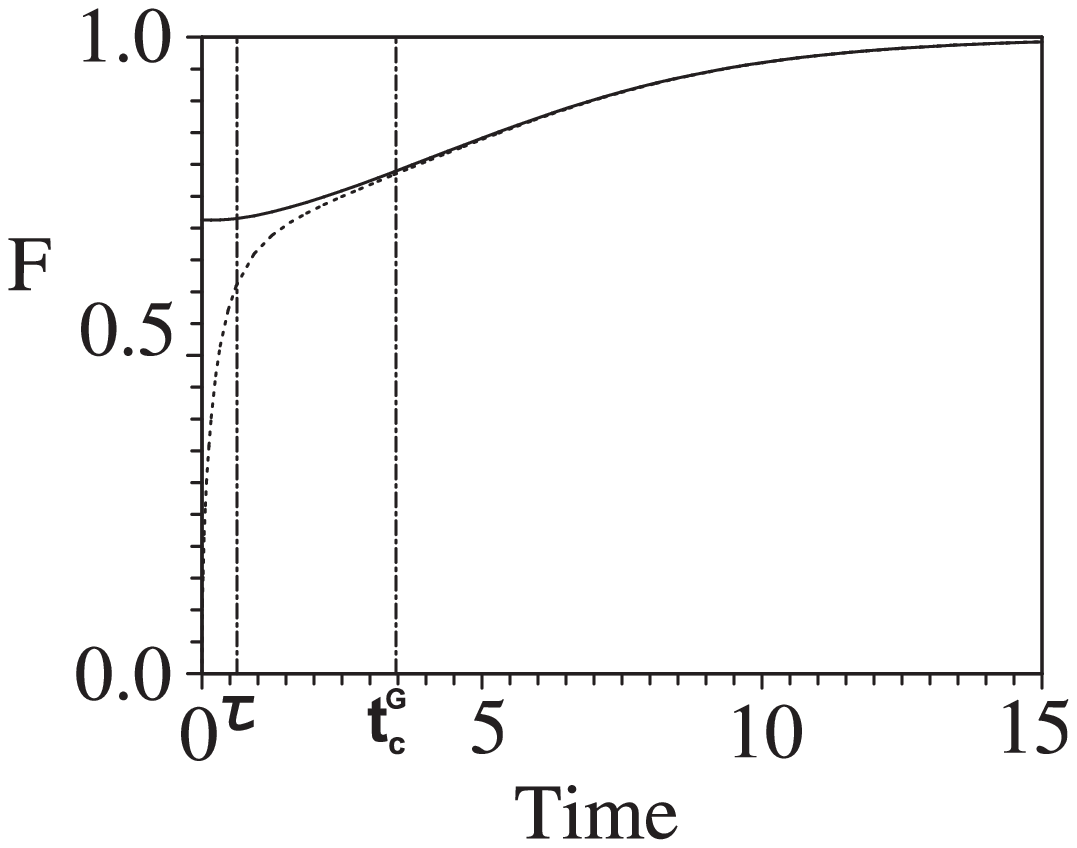}
\end{center}
\caption{Fidelity between the coherent superposition states and
the GSS. The parameters used: $\beta_0 = 2.0$, $\omega=1$,
$k=0.1$, $r_0 =1$, $\alpha_0 \simeq 0.0$, $\phi_0 = 0$, $n_B = 0$,
$\nu_0 = 0$, $\theta = 0$ (solid line) and $\theta = \pi$ (dashed
line).} \label{fidelidade2}
\end{figure}

\begin{figure} [!h]
\begin{center}
\includegraphics [scale=0.6] {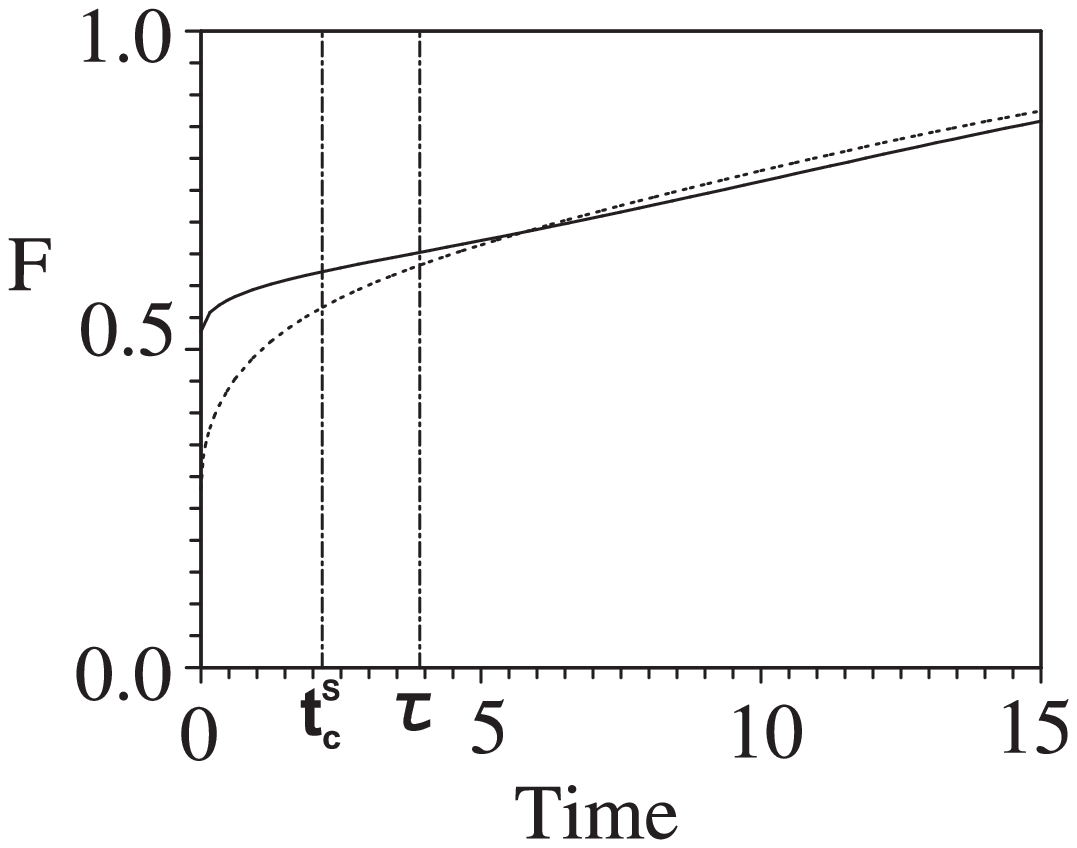}
\end{center}
\caption{Fidelity between the coherent superposition states and
the GSS. The parameters used: $\beta_0 = 0.8$, $\omega=1$,
$k=0.1$, $r_0 =1$, $\alpha_0 \simeq 0.0$, $\phi_0 = 0$, $n_B = 0$,
$\nu_0 = 3$, $\theta = 0$ (dashed line) and $\theta = \pi$ (solid
line).} \label{fidelidade3}
\end{figure}

\begin{figure} [!h]
\begin{center}
\includegraphics [scale=0.6] {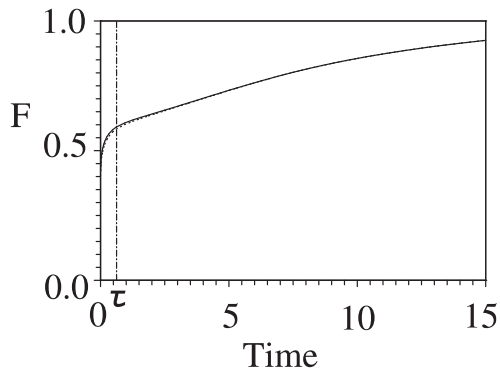}
\end{center}
\caption{Fidelity between the coherent superposition states and
the GSS. The parameters used: $\beta_0 = 2.0$, $\omega=1$,
$k=0.1$, $r_0 =1$, $\alpha_0 \simeq 0.0$, $\phi_0 = 0$, $n_B = 0$,
$\nu_0 = 3$, $\theta = 0$ (dashed line) and $\theta = \pi$ (solid
line).} \label{fidelidade4}
\end{figure}

To understand the (big and cumbersome) expressions for the
fidelity we plot the fidelity against time in figures
\eqref{fidelidade1} - \eqref{fidelidade4} (the time scale of the
graphics is: $\mbox{time} = k t$). The choice $\alpha_0 \simeq
0.0$ is due to numeric computations. In fact we use $\alpha_0 =
10^{-20}$. In each figure, we show the decoherence time $\tau$ for
the CSS, {\it i.e.} $\tau = (4 |\beta_0|^2 k)^{-1}$ and the
squeezing characteristic time for CSS and for GSS -- equations
\eqref{tccat} and \eqref{tcgauss} -- in the vertical lines.

Analyzing the figures we can conclude the same as we did before,
but in a more quantitative way. In figures \eqref{fidelidade1} and
\eqref{fidelidade2} we plot the fidelity between a squeezed
Gaussian state with $\nu_0 = 0$ and even and odd superposition
states with $\beta_0 = 0.8$ and $\beta_0 = 2.0$, respectively. One
can see that the initial fidelity is always high for the even
superposition state with short values of $\beta_0$, while when we
increase this parameter the fidelity decreases. The even
superposition with $\beta_0 = 0.8$ is approximately a Gaussian
state -- apart the fact that it has negative Wigner function --
and it behaves like the GSS, i. e., the squeezing property
dominates the other quantum ones. When we increase $\beta_0$, the
interference property dominates and the state does not resemble
the GSS. For values of $\beta_0$ not respecting the inequality
\eqref{catcondition} we see that the even superposition state does
not possess high fidelity and rapidly behaves like the odd one.

In figures \eqref{fidelidade3} and \eqref{fidelidade4} we show the
results for the GSS with $\nu_0 = 3$ and even and odd
superposition states with $\beta_0 = 0.8$ and $\beta_0 = 2.0$,
respectively. Both the even and the odd superposition states
behave similarly in these cases. In the first case the GSS is not
``so squeezed'' as the even superposition state and the fidelity
between them is short. In the second case, the interference
property dominates the superposition states and their fidelity
against the GSS behaves practically equal.
\\

\section{Summary and Conclusions} \label{conclusions}

In this work we study in detail, quantum properties of coherent
superposition states (even and odd superposition states) and of
displaced, squeezed, thermal states. We analyze the squeezing (via
the Wigner function and the covariance matrix determinant), the
oscillations in photon distribution (through the diagonal term of
the density operator $\rho_{n n} = P_n$) and the von Neumann
entropy ($S[\rho] = - \tr [\rho \ln \rho])$ for both cases. We
show that in superposition states each property has different
characteristic time (being the ``squeeze time'' lesser than the
decoherence time) while in the GSS we found only one
characteristic time \eqref{tcgauss}. The even superposition state
shows squeezing, just like the GSS, and we can access this
property via the covariance matrix. We show that, for both
cases (even superposition state and GSS) the squeezing effect only
can be observed in special initial conditions -- equations
\eqref{catcondition} and \eqref{inequal} -- and in experimentally
 accessible time scales -- equations \eqref{tccat} and \eqref{tcgauss}. Since it is ``easy'' to
observe quadratures (for example with homodine detection), one can
use the squeezing effect to study quantum information, provided
the initial conditions fulfill the inequalities cited before.
Finally, we compare these effects more quantitatively by analyzing
the fidelity between the GSS and the superposition states (even
and odd), concluding the same, i. e., if the states respect the
inequalities mentioned before (the GSS and the even superposition
state for instance), the fidelity has high value.

\textbf{Acknowledgments.} The authors thank funding from Brazilian agencies CNPq and FAPEMIG.

\end{document}